The Proposed Silicate-Sulfuric Acid Process: Mineral Processing for In Situ Resource Utilization (ISRU)

Seamus L. Anderson*[1], Eleanor K. Sansom[1], Patrick M. Shober[1], Benjamin A.D. Hartig[1], Hadrien A.R. Devillepoix[1], Martin Towner[1]

[1]Space Science and Technology Center, Curtin University, GPO Box U1987, Perth, WA 6845, Australia
*Corresponding author: seamus.anderson@postgrad.curtin.edu.au

**Abstract**

Volatile elements and compounds found in extra-terrestrial environments are often the target of In Situ Resource Utilization (ISRU) studies. Although water and hydroxide are most commonly the focus of these studies as they can be used for propellant and human consumption; we instead focus on the possible exploitation of sulfur and how it could be utilized to produce building materials on the Moon, Mars and Asteroids. We describe the physical and chemical pathways for extracting sulfur from native sulfide minerals, manufacturing sulfuric acid in situ, and using the produced acid to dissolve native silicate minerals. The final products of this process, which we call the Silicate-Sulfuric Acid Process (SSAP), include iron metal, silica, oxygen and metal oxides, all of which are crucial in the scope of a sustainable, space-based economy. Although our proposed methodology requires an initial investment of water, oxygen, and carbon dioxide, we show that all of these volatiles are recovered and reused in order to repeat the process. We calculate the product yield from this process if it were enacted on the lunar highlands, lunar mare, Mars, as well as an array of asteroid types.

## 1. Introduction

Humanity's renewed interest in deep-space exploration will bring to bear countless challenges we as a species have not yet faced. As humans plan to depart from Earth for longer durations and further distances than ever before, they must be equipped with an increasingly large stockpile of resources in order to survive and thrive on their voyages. An alternative to this 'bring everything' approach, is to instead make use of the resources present at the various destinations, such as the Moon and Mars. This concept is known as in situ resource utilization (ISRU), and although the basic principle is not a novel idea [1], its numerous demonstrations and theoretical implementations are emerging at a faster rate than ever before [2,3,4,5].

Previous ISRU studies have investigated a wide range of possibilities for extracting and utilizing materials from celestial bodies across the solar system, from rare earth elements on certain asteroids [6], to the constructional uses of regolith for radiation shielding on the Moon and Mars [7]. Much of this current literature regarding ISRU focuses on extracting volatile elements, chiefly oxygen and hydrogen, which are often locked away in the form of non-volatile minerals. The importance of these two resources cannot be understated, since they can be used not only as breathable air and potable water for human consumption but could also serve as propellant for rocket engines [5,8,9]. To that end, carbon also plays an important role in martian ISRU, since it can be combined with hydrogen to make fuel for methane-based rocket engines [10,11,12]. Extracting volatile elements and refining propellants on the surfaces of the Moon and Mars will significantly reduce the mass required to launch from Earth, and therefore the cost of spacecraft for interplanetary missions. That being said, we focus on an often overlooked volatile element that is present on the Moon, Mars and numerous asteroids: sulfur.

A study by Vaniman et al. [13] illustrated possible uses for lunar sulfur, from sealants to electricity generation and storage. A series of experiments have also evaluated the mechanical properties and durability of concrete made from lunar soil simulant and elemental sulfur [14,15]. Unfortunately the deterioration of such concretes due to simulated lunar thermal cycling and degassing under vacuum has reduced the interest in this application in recent years.

Here, we will instead explore the physical and chemical pathways for manufacturing sulfuric acid ($H_2SO_4$) from native sulfur in order to dissolve silicate minerals, also native to these bodies. Doing so would produce considerable amounts of iron metal, silica, oxygen as well as other useful building materials for permanent human settlement in space. We call this methodology: The Silicate-Sulfuric Acid Process (SSAP). Although mineral processing for ISRU using sulfuric acid has been examined in the past [16, 17], harvesting oxygen from the less abundant mineral ilmenite ($FeTiO_3$) was the main focus. In this paper, a variety of highly-abundant silicate minerals are the principle focus for refinement into building materials.

Using a relatively small investment of other volatile elements including carbon, oxygen, and hydrogen, this process could be carried out on the Moon, Mars, and many Asteroids. Although this initial investment could be costly, many of these elements can be found in situ especially for mission profiles where the primary goal is to extract water, such as on C-type asteroids, the lunar poles, and high-latitude locations on Mars. The other main invested volatile, carbon, is also found on C-type asteroids, as well as on the surface and atmosphere of Mars. Furthermore, we will show that the SSAP allows for inherent recycling of the invested volatile elements, for continuous reuse.

### 1.1 Sulfur Availability

In the ordinary and carbonaceous chondrite meteorite groups, which originated from S and C-type asteroids, respectively [18,19,20], sulfur is fairly abundant (~2.5 wt%). In ordinary chondrites it is almost entirely contained in the mineral troilite (FeS) [21], while carbonaceous chondrites usually contain pyrrhotite ($Fe_{1-0.8}S$) and pentalandite (($Fe, Ni)_9S_8$) instead [22]. Martian dust can contain 2.5 wt % sulfur [23], while sulfates (e.g. $MgSO_4$, $FeSO_4$) are regularly detected in martian soils [24,25]. On the Moon, troilite exists across most of the surface, although in comparatively low abundance (<0.5 wt %) [26]. In the permanently shadowed crater regions near the poles, the sulfur content could be as high as 1 wt% in the form of $SO_2$ ice [27]. Although sulfur may not be one of the most abundant elements on these bodies, its presence as a minor element still offers an opportunity to utilize it.

### 1.2 Silicate-Bound Resources

Silicate minerals are an obvious target for acid-driven dissolution, since they are abundant on essentially all terrestrial bodies. They usually exist in the general formula: $α_iSi_jO_k$; where α can represent Mg, Fe, Ca, Al, Na, K, or other metals present within the crystal lattice. On the Moon, silicate minerals mostly consist of plagioclase ($CaAl_2Si_2O_8$), pyroxene ($[Mg,Fe]_2Si_2O_6$) and some olivine ($[Mg,Fe]_2SiO_4$), totaling more than 70 vol% of the regolith [28]. Similarly, most soils on Mars have a total silicate abundance near 80 wt%, where it is also mostly composed of plagioclase, pyroxene and olivine [29]. On C-type asteroids, with similar mineralogies to CM and CI meteorites, silicates in the form of olivine and water-bearing phyllosilicates, collectively account for between 75 and 95 % of the total weight [22]. Meanwhile S-type asteroids contain mostly olivine and pyroxene with some plagioclase, totaling between 75 and 90 wt% of these bodies [30, 19]. Once the metal components in these minerals are liberated they would be extremely valuable and useful for building large structures and other essential hardware off-world.

Iron is a prime example since it is already widely used for building large structures here on Earth, particularly when it is combined with other elements to form steel. Currently, the only way to obtain steel off-world is to launch it into orbit from Earth, a very expensive means of construction. Alternatively, producing iron metal at the desired destination could significantly reduce mission costs when the goal is to make large, permanent structures and equipment in space.

Silica ($SiO_2$), another main product of the SSAP, can serve as the precursor for fused quartz, which has been used as spacecraft windows on the Space Shuttle orbiters and the International Space Station [31]. Fused quartz is made by melting silica grains (~1650 °C) either under vacuum or in an inert atmosphere. Due to this high melting temperature, fused quartz can be used in some high temperature environments. This additionally makes it a possible candidate for constructing some of the equipment required for the processes we describe below.

As previously discussed, oxygen has obvious applications for human space exploration, since it can be used as breathable air. When carefully combined with hydrogen it can also be used as rocket propellant; otherwise it forms pure water for both human consumption and industrial processes.

The various oxides formed by the SSAP have an extensive variety of niche uses, possibly the most useful of which is the molten electrolysis of aluminum oxide to form pure metal. Although this refinement is not a focus of this paper, aluminum metal could feed into the fabrication of lightweight, high-strength alloys.

## 2. Physical and Chemical Pathways of the SSAP

The proposed Silicate-Sulfuric Acid Process consists of four main stages. The first stage contains optional pre-processing steps so that the SSAP can be compatible with other resource extraction methods such as water harvesting. The second stage entails the synthesis of sulfuric acid, either from indigenous minerals or from recycled sulfur, water and oxygen. The third stage then uses this sulfuric acid to dissociate the silicate minerals into silica and sulfate minerals; the former of which is a final product and is removed from further processing. The sulfates are then thermally decomposed, and some of their products reduced to form metal and simple oxides in the fourth and final stage. Each of these stages is discussed below in further detail. Fig. 1 provides an overview of the SSAP in the form of a flow diagram.

For each of the reaction steps, we calculate the change in Gibbs free energy (ΔG), at 20 °C and 1 bar, from reported values in the NIST Standard Reference Database [32]. Eq. (1) shows this relation to enthalpy (H), entropy (S) and temperature (T). For the minerals that did not appear in this database, we compiled their thermodynamic properties from individual sources (listed in supplementary materials). Although this approach of combining multiple databases for comparison and calculation is not ideal, it does provide an estimate for the energy requirements at each step.

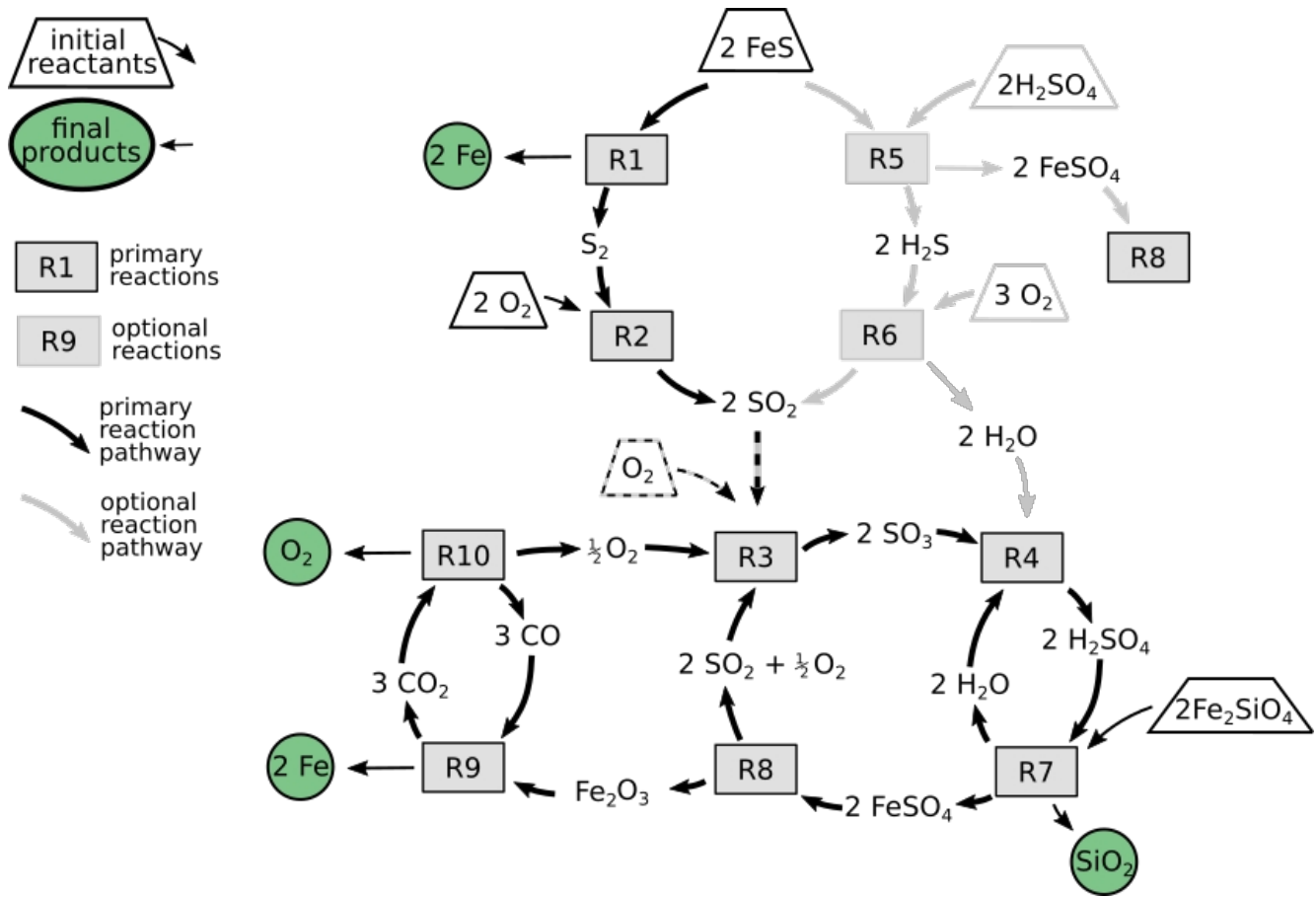

Fig. 1. This flowchart illustrates the chemical reaction pathways for the SSAP, while also highlighting the recycling of oxygen, sulfur dioxide, carbon monoxide and water. For simplicity, this shows only the processing of fayalite ($Fe_2SiO_4$). Final products are signified by a green circle, input reactants are outlined by trapezoids, and named reactions steps (see text) appear in rectangles. The upper half of this figure shows two pathways for processing troilite, either through thermal decomposition (left) or acid-dissociation (right) when sulfuric acid is available.

$$\Delta G = \Delta H - T\Delta S \qquad (1)$$

We also calculate Gibbs free energy at pressures of $10^{-2}$, $10^{-5}$, and $10^{-8}$ bar ($G_p$) for each compound we assess, using Eq. (2), where R is the ideal gas constant, n is number of moles of gas, and P is pressure. From these Gp values, we also calculate the temperature-dependent equilibrium constant ($K_{eq}$) at those pressures via Eq. (3). This enables us to predict the equilibrium composition of each reaction using Eq. (4) and (5), where a compound's concentration is signified by square brackets (e.g. [A]). This is especially important for reactions that involve thermal decomposition, as we will show that performing these steps under vacuum can significantly reduce the temperatures required for the reaction to proceed.

$$G_p = G + nRT\ln(P) \quad (2)$$

$$K_{eq} = e^{\left(\frac{-\Delta Gp}{RT}\right)} \quad (3)$$

$$K_{eq} = \frac{[X]^x [Y]^y}{[A]^a [B]^b} \quad (4)$$

$$aA + bB = xX + yY \quad (5)$$

### 2.1 Pre-processing

Ideally the first step in the SSAP consists of mechanically crushing larger silicate rocks to reduce the average particle size, which would allow for a quicker reaction with the sulfuric acid. This can be bypassed if regolith is used instead of large boulders or rocks. The bulk material should also be magnetically separated to extract native Fe-Ni metal grains present on the Moon and some asteroids [21, 28]. Although these metal grains could be processed along with the silicates, doing so would be redundant as iron metal is an end product of the SSAP. Magnetic separation on the Moon could also separate the weakly magnetic mineral ilmenite from the silicates, which can separately be reduced into iron metal, titanium oxide and oxygen [33,34,35].

As we discussed earlier, water is possibly the most valuable resource to be harvested off-world. To avoid further complicating its extraction, water-bearing minerals and regoliths can be heated, and their evolved vapors collected in a cold trap, prior to the rest of the material being subjected to the SSAP. On the Moon, this means that water-ice rich regoliths [36] should be heated to 150 °C in order to sublimate the ice into vapor [37]. For C-type asteroids, the phyllosilicates should be heated to liberate the lattice-bound $OH^-$ molecules [38], which will also recrystallize much of the phyllosilicates into olivine [39]. Alternatively, if sulfuric acid were applied to water-bearing ores prior to heating, the silicates would still dissociate but the aqueous solution will be more dilute, which would likely slow the reaction rate.

### 2.2 Sulfuric Acid Synthesis

The sulfuric acid synthesis stage of the SSAP consists of thermal decomposition of sulfide minerals, followed by the industry-proven wet sulfuric acid process (WSA) [40], which creates sulfuric acid from sulfur-bearing gases. The main ore for sulfur on the Moon and S-type asteroids comes in the form of troilite (FeS), where as for C-type asteroids, the slightly more sulfur-rich pyrrhotite ($Fe_{(1-0.8)}S$) is the dominant sulfide mineral [22]. For simplicity in our description we assume that the sulfides appear in their stoichiometric flavor troilite (FeS). For Mars, the main sulfur-ore consists of sulfates (e.g. $MgSO_4$, $FeSO_4$) which can be heated to release $SO_2$ gas and form solid oxides (discussed later in Reaction 8).

$$2\ FeS\ (s) + 284\ kJ\ mol^{-1} \rightarrow 2\ Fe\ (s) + S_2\ (g) \quad (R1)$$
$$S_2\ (g) + 2\ O_2\ (g) \rightarrow 2\ SO_2\ (g) + 680\ kJ\ mol^{-1} \quad (R2)$$

$$2\,SO_2\,(g) + O_2\,(g) \quad \rightarrow \quad 2\,SO_3\,(g) \quad + 142 \text{ kJ mol}^{-1} \quad (R3)$$
$$2\,H_2O\,(l) + 2\,SO_3\,(g) \quad \rightarrow \quad 2\,H_2SO_4\,(aq) \,+\, 90 \text{ kJ mol}^{-1} \quad (R4)$$

Reaction 1 shows the thermal decomposition of troilite and yields both iron metal and sulfur gas by heating to approximately 1250 °C in vacuum. More sulfur-rich minerals such as pyrrhotite and pyrite will begin releasing their sulfur component before this temperature [41]. This decomposition has been explored in meteorite heating experiments [38] which show that CM chondrites undergo a minor sulfur volatilization event around 550°C, followed by a major outgassing event at 1200 °C. In these experiments, the sulfide minerals were not separated from the rest of the meteorite sample when they were heated and formed $SO_2$ gas rather than pure $S_2$. It is unclear how much, if any of the iron in the sulfides was oxidized into $Fe_2O_3$ or $Fe_3O_4$ in this more oxygen-available environment. If the sulfide minerals were instead isolated from the rest of the bulk material before they were heated, the resultant gas would more likely be comprised of pure sulfur, while the iron within the sulfides would not likely be oxidized, resulting in pure iron metal. This iron metal is the first product harvested from the SSAP.

If pure sulfur gas is produced in Reaction 1, it must be exothermically combined with oxygen to yield sulfur dioxide (Reaction 2). If the thermal decomposition of the sulfide minerals results instead in sulfur dioxide, then Reaction 2 can be bypassed. The sulfur dioxide is then subjected to the WSA process (Reactions 3-4), whereby the gas is oxidized in the presence of a vanadium oxide catalyst between 400 and 620 °C (Reaction 3), in order to form sulfur trioxide. It is important to note that this catalyst is not depleted during the reaction and can be reused. The resulting sulfur trioxide is then exothermically hydrated before being condensed to form highly concentrated sulfuric acid in Reaction 4. Although Reactions 2-4 require an investment of oxygen and hydrogen, we will show in later stages that they will be recovered and can be reused, such that no volatiles are lost or wasted.

Alternatively, pre-existing sulfuric acid either brought to location or created in earlier processing, can be reacted with troilite to produce iron sulfate and hydrogen sulfide gas (Reaction 5). The resulting gas can then be burned with oxygen to form water and sulfur dioxide (Reaction 6), both of which are used in the above reactions to produce sulfuric acid. The processing of the iron sulfate in Reaction 5, will be elaborated on further in the next subsection. This alternative approach may be more logistically feasible since the sulfide and silicate minerals would not need to be separated prior to their reaction with the acid. This approach would also be ideal on the martian surface, as the main sulfuric ores are various sulfates [29] that should dissolve in sulfuric acid.

$$FeS\,(s) + H_2SO_4\,(aq) \quad \rightarrow \quad FeSO_4\,(s) + H_2S\,(g) \quad + 103 \text{ kJ mol}^{-1} \,(R5)$$
$$H_2S\,(g) + {}^3\!/_2\,O_2\,(g) \quad \rightarrow \quad H_2O\,(g) + SO_2\,(g) \quad + 504 \text{ kJ mol}^{-1} \,(R6)$$

### 2.3 Silicate Dissolution and Silica Extraction

Once the sulfuric acid is synthesized, the SSAP proceeds to the next stage: silicate dissolution. By combining the acid with the silicate minerals (Reaction 7), this stage produces silica, water, and sulfate minerals; the last of which will be broken down further in final stage of the SSAP. The reactions between the silicates and the acid are listed in Table 1.

$$\alpha_i Si_j O_k\,(s) + H_2SO_4\,(aq) \rightarrow \alpha_m SO_4\,(aq/s) + H_2O\,(l) + SiO_2\,(s) \quad (R7)$$

Table 1. The generalized reactions between sulfuric acid and the end-members of each silicate mineral. Change in Gibbs free energy was calculated for 20 ºC at 1 bar. The silica and sulfate products will both precipitate and be dissolved in solution depending the conditions of the reaction chamber. The negative values for ΔG in this table indicate that each reaction is exothermic and will proceed under standard temperature and pressure.

| Reactants | | | | | Products | | | |
|---|---|---|---|---|---|---|---|---|
| Silicates (s) | | | Sulfuric Acid (aq) | | Water (l) | Silica (s/aq) | Sulfates (s/aq) | ΔGº [kJ mol$^{-1}$] (20 °C, 1 bar) |
| Mineral | Endmember | Formula | | | | | | |
| Olivine | Fayalite | $Fe_2SiO_4$ | 2 $H_2SO_4$ | → | 2 $H_2O$ | $SiO_2$ | 2 $FeSO_4$ | -292 |
| | Forsterite | $Mg_2SiO_4$ | 2 $H_2SO_4$ | | 2 $H_2O$ | $SiO_2$ | 2 $MgSO_4$ | -258 |
| Pyroxene | Ferrosilite | $Fe_2Si_2O_6$ | 2 $H_2SO_4$ | | 2 $H_2O$ | 2 $SiO_2$ | 2 $FeSO_4$ | -1220 |
| | Enstatite | $Mg_2Si_2O_6$ | 2 $H_2SO_4$ | | 2 $H_2O$ | 2 $SiO_2$ | 2 $MgSO_4$ | -270 |
| | Wollastonite | $Ca_2Si_2O_6$ | 2 $H_2SO_4$ | | 2 $H_2O$ | 2 $SiO_2$ | 2 $CaSO_4$ | -2080 |
| Plagioclase | Anorthite | $CaAl_2Si_2O_8$ | 4 $H_2SO_4$ | | 4 $H_2O$ | 2 $SiO_2$ | $CaSO_4$, $Al_2(SO_4)_3$ | -2250 |

     Although we list the pure end-members of these minerals, nearly every silicate grain native to terrestrial bodies is actually a solid solution, with a varying proportion of the appropriate metal cation coexisting in the same crystal lattice. It is for simplicity that we examine the reactions of the pure end members with sulfuric acid.

     A suite of previous experiments describe in detail, the acid-silicate reactions listed in Table 1 [42,43,44,45,46]. The results of these experiments show the general trend that the silicate minerals are broken down into a hydrated amorphous silica gel, while the cations (Fe, Mg, etc.) are released into the water-acid solution and eventually precipitate into their hydrated sulfate counterparts. Minor amounts of iron oxides also form from olivine when the initial aqueous sulfuric acid solution is less concentrated [43]. The mixture should be mechanically perturbed or mixed to prevent a nonreactive product layer to form on the surface of unreacted silicate grains. Since previous experiments did not mix or perturb the rock-acid mixtures, it is unclear exactly how quickly this step will progress.

     Once all of the initial silicates have reacted with the acid, the fluid can be evaporated such that any excess water or acid can be collected for later use. The evaporation will cause the ions in the solution to precipitate into the sulfates listed in Table 1. At this point most of the solid products will still likely contain water and can be dehydrated by heating to 100 °C under vacuum. This released water can be immediately reused in Reaction 4. This step of separating water and unreacted acid from dissolved components is one of the key factors that will determine the overall efficiency of the entire SSAP in terms of energy. Adding more acid-water solution at the beginning of Reaction 7 will cause it to progress more quickly; however this also requires more energy to evaporate the remaining liquid. In later sections we estimate discuss efficiency bottlenecks for the SSAP.

## 2.4 Metal and Oxide Production

The final stage of the SSAP produces iron metal, oxygen, as well as metal oxides, via thermal decomposition and carbothermal reduction. The sulfates previously produced in Reaction 7 (Table 1) can be intermixed for this next step, since calcium, aluminum, magnesium and iron(II) sulfate each have distinct thermal decomposition temperatures. Heating the sulfates will decompose the iron(II) sulfate into iron(III) oxide ($Fe_2O_3$), sulfur dioxide, and oxygen, as shown in Reaction 8 (Table 2). Simultaneously, the aluminum sulfates will decompose into aluminum oxide, sulfur dioxide and oxygen. Since iron(III) oxide is ferromagnetic, while aluminum oxide, calcium and magnesium sulfate are diamagnetically susceptible, the iron(III) oxide can be magnetically separated before further heating. The remaining magnesium and calcium sulfates will thermally decompose into their corresponding metal oxides at higher temperatures. The equilibrium compositions for each of these reactions is shown in Fig. 2. Further refining these oxides into pure metals (Al, Mg, Ca) is not addressed in this work; we will only describe the extraction of iron metal from iron(III) oxide.

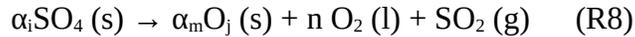

$$\alpha_i SO_4 \,(s) \rightarrow \alpha_m O_j \,(s) + n\, O_2 \,(l) + SO_2 \,(g) \quad (R8)$$

Table 2. Generalized thermal decomposition reactions for produced sulfates. The Oxides described here are in their simple form (MgO:Magnesia, $Al_2O_3$:Alumina). The positive values attained for ΔG° indicate that heat-energy is required for the reaction to proceed (see Fig. 2). These reactions are collectively referred to as Reaction 8.

| Reactants | | Products | | | |
|---|---|---|---|---|---|
| Sulfate (s) | | Sulfur Dioxide (g) | Oxygen (g) | Oxide (s) | ΔG° [kj/mol] (20 °C, 1 bar) |
| 2 $FeSO_4$ | → | 2 $SO_2$ | ½ $O_2$ | $Fe_2O_3$ | 306 |
| 2 $MgSO_4$ | | 2 $SO_2$ | $O_2$ | MgO | 557 |
| 2 $CaSO_4$ | | 2 $SO_2$ | $O_2$ | CaO | 868 |
| $Al_2(SO_4)_3$ | | 3 $SO_2$ | $^3/_2\, O_2$ | $Al_2O_3$ | 621 |

The gases produced in Table 2 can be collected and reused in Reactions 2 and 3 to produce more sulfuric acid. Although not all the invested oxygen is recovered from iron(III) oxide the final steps will net a 1 mole surplus of $O_2$.

Iron metal is produced by reducing iron(III) oxide using carbon monoxide in Reaction 9. Although Chen et al. [47] show that this process consists of multiple steps: $Fe_2O_3 \rightarrow Fe_3O_4 \rightarrow FeO \rightarrow Fe$, oxidizing the ambient CO atmosphere all along the way, we forgo these intermediates and represent the reaction more concisely.

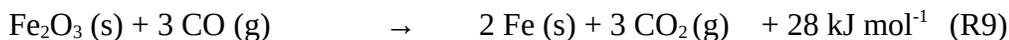

$Fe_2O_3$ (s) + 3 CO (g) $\quad\rightarrow\quad$ 2 Fe (s) + 3 $CO_2$ (g)  + 28 kJ mol$^{-1}$  (R9)

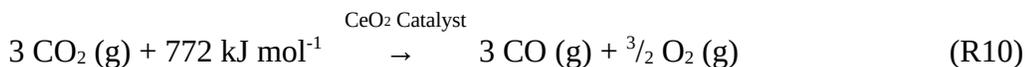

$$3\, CO_2 \,(g) + 772 \text{ kJ mol}^{-1} \xrightarrow{CeO_2 \text{ Catalyst}} 3\, CO\,(g) + {}^3/_2\, O_2\,(g) \quad (R10)$$

The final step of the SSAP is to recover the oxygen locked away in the carbon dioxide at the end of Reaction 9, while also replenishing the supply of carbon monoxide for repeating the very same reaction. This is done by electrolyzing $CO_2$ with a cerium oxide catalyst near 500 °C [48], shown in Reaction 10. We would like to note that the energy required that we list in Reaction 10 is likely an overestimate since we do not consider the effects of the catalyst in our calculations. Now that the CO gas has been replenished for repeat use, 1 mole of $O_2$ has also been created for every 2 moles of iron metal produced, and the SSAP is complete.

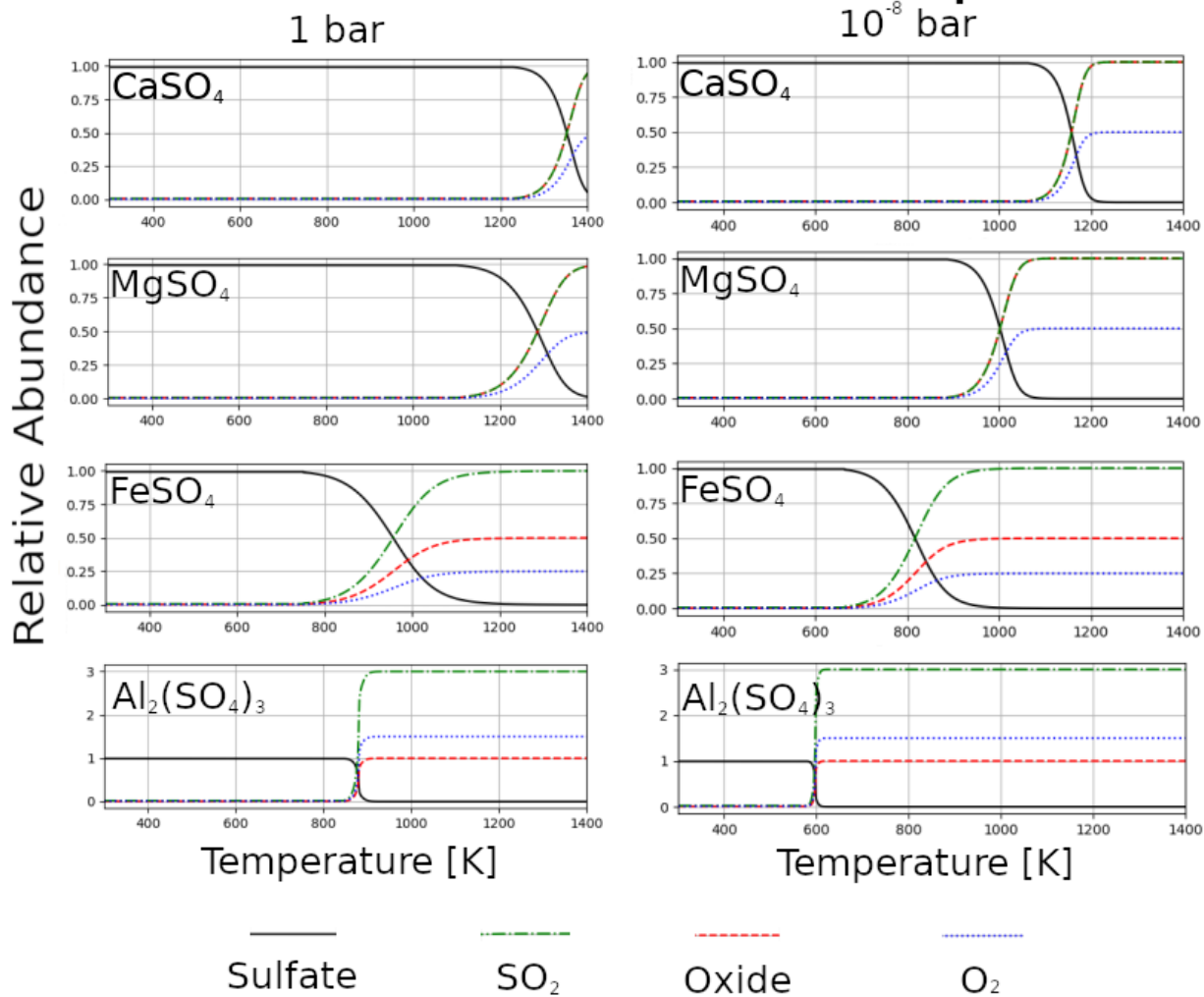

Fig. 2. The calculated equilibrium compositions for each major sulfate species at 1 bar (left column) and $10^{-8}$ bar (right column). This shows the general trend that reducing the ambient pressure lowers the temperatures required for the reactions to proceed. Note the legend at the bottom of the figure. The Y axes represent molar abundance of each compound.

## 3. Results and Discussion

We have calculated the theoretical yield of the SSAP for each terrestrial body listed in Table 3. We assume an initial silicate mass of one ton, with mineral chemistry and abundance representative of each particular body. By taking the product of each mineral's abundance and its end member molar percentage, we calculated the total number of moles of each silicate end member. With this, we use Reactions 7-10 to calculate how many moles, and by extension kilograms, of each resource could be produced from the SSAP. For H, L, and LL chondrites (S-type asteroids) we obtained average mineral abundances from [30], and mineral compositions from [49,50,51]. For C-type asteroids we used mineral data reported by [22] for the meteorites: Murchison, Orgueil and Allende (CM, CI, CV chondritesrespectively). The C and S-type asteroid calculations include contributions from native sulfides. For the lunar highlands and mare calculations, we averaged the bulk chemical compositions reported in [28] including the ilmenite and sulfide contributions. This calculation assumes that the ilmenite cannot be separated by pre-processing and is included in the silicate dissolution step (Reaction 7). For the martian calculations we averaged the values reported in [29] across both models and localities (Gusev and Meridiani), but we did not include contributions from native sulfates, which would increase the total yield of iron metal and oxygen.

Table 3. Theoretical yield for various solar system bodies, assuming 100% efficient processing of 1000 kg of native silicates. The Oxides described here are in their simple form (*e.g.* MgO:Magnesia, $Al_2O_3$:Alumina, etc.)

| Body | Recoverable Resource [kg] | | | | | | | | Reference |
|---|---|---|---|---|---|---|---|---|---|
| Luna | Fe (metal) | $O_2$ | $SiO_2$ | $Al_2O_3$ | MgO | CaO | $H_2O$ | $TiO_2$ | |
| Mare | 143 | 31 | 492 | 143 | 100 | 123 | - | 46 | [28] |
| Highlands | 53 | 13 | 475 | 237 | 85 | 142 | - | 8 | [28] |
| **Mars** | 121 | 35 | 514 | 124 | 77 | 96 | - | - | [29] |
| **Asteroids** | | | | | | | | | |
| C-type (CI-Orgueil mineralogy) | 65 | 3 | 430 | - | 436 | - | 118 | - | [22] |
| C-type (CM-Murchison mineralogy) | 375 | 125 | 260 | - | 174 | - | 87 | - | [22] |
| C-type (CV-Allende mineralogy) | 267 | 53 | 393 | 4 | 359 | 2 | - | - | [22] |
| S-type (H | 132 | 27 | 509 | 25 | 323 | 20 | - | - | [30,49,50,51] |
| Chondrite mineralogy) | | | | | | | | | |
| S-type (L Chondrite mineralogy) | 170 | 35 | 490 | 22 | 308 | 20 | - | - | [30,49,50,51] |
| S-type (LL Chondrite mineralogy) | 164 | 40 | 473 | 21 | 304 | 18 | - | - | [30,49,50,51] |

### 3.1 Products and Uses

Our calculations show that a significant quantity of building material can be obtained on the Moon, Mars, and asteroids (Fig. 2). Table 3 shows that CM-like C-type asteroids are the most fruitful candidate for the SSAP, as it would produce the most iron metal and oxygen, with harvested water being an added benefit. Resource utilization on S-type asteroids would particularly benefit from the SSAP, as they would be otherwise considered relatively resource-poor, due to their lack of native water, while also hosting a relatively high abundance of sulfur. For some C-type asteroids, we list water as a SSAP-product only because their silicates contain significant native water that is incidentally released, otherwise water is not a product of the SSAP. For lunar operations, the lunar mare would be preferable to the lunar highlands in terms of a more useful product yield, since it is more highly concentrated in iron which can be used in 3D printing (discussed below). As for the longevity of SSAP operations at a destination, Mars may be the best candidate since it has abundant water ice as well as the highest relative abundances of carbon and sulfur. This would allow for higher tolerances in volatile loss.

On the lunar surface, the SSAP could considerably contribute to the Artemis program's goals of establishing a sustainable presence on the surface on the Moon, by providing some building materials in situ. The produced iron metal will likely take the form of small particles, which can be used as the feedstock for direct metal laser sintering to 3D print components or structures including landing pads, radiation shields, or electrical wire. The silica produced here will also likely be in granular form, which can be melted in a cast to produce windows for future habitats. Alumina is a natural insulator of electricity, making it ideal for encasing power cables for extra-terrestrial solar power plants. As we mentioned earlier, alumina could be further reduced into aluminum metal, although this pathway is beyond the scope of the SSAP.

We envision a logistical framework, whereby a cargo spacecraft could deliver to a planetary surface: a metal 3D printer, a casting furnace, and a SSAP-refinery including some initial sulfuric acid, water, and carbon dioxide. The refinery could begin producing silica for the furnace and iron for the printer, to construct the hull or main body of a habitat in situ. Later cargo missions can deliver robotic workers and more specialized components such as airlock doors or life support systems to be installed. Although the mission architecture for the Artemis program prescribes an Earth-fabricated habitat, any attempts at permanent human settlement will likely require a process to create infrastructure in situ.

## 3.2 Engineering and Logistical Considerations

Although most of the steps within our proposed process are supported by a suite of previous experiments including some industry-proven methodologies, its effectiveness should be validated by performing these techniques on lunar, martian, and asteroid regolith simulants. These experiments will help to determine reaction rates and will characterize some of the engineering challenges that will inevitably become apparent.

The first experimental validation should be focused on the feasibility of processing the bulk material, not solely the silicates, for a given locale. For instance, the magnetic force required for collecting native iron metal and iron oxide must be determined. This is especially important for martian regolith, which contains considerable iron oxide. Once the separation is complete, the remaining non-magnetic material should be dissolved in acid (Reaction 7) to determine what problems, if any, might arise from insoluble impurities while also characterizing the reaction rate. Previous experiments [46] also indicate that Reaction 7 progresses more quickly if the temperature is slightly elevated. Keeping Reaction 7 in thermal contact with Reaction 8 (thermal decomposition) may be an efficient way to conserve energy.

The overall efficiency of the SSAP will be heavily influenced by the efficiency of Reaction 7. The less water and mechanical perturbations required to fully react the silicates (or non-magnetic material) will reduce the overall energy requirements for this step. Unfortunately we cannot accurately calculate the total energy required to enact the SSAP from start to finish, as it is unclear which hydrated sulfates would form (monohydrate, pentahydrate, etc.) in Reaction 7. Additionally, an accurate calculation would require knowledge of the power and duration needed to operate the vacuum systems in Reaction 8. These reasons underline the need for experimental investigation in the future.

As we briefly mentioned in the previous subsection, the long term viability of the SSAP will also depend on how much of the volatile compounds can be retained, especially when trapping evolved gases in Reaction 8. Having the ability to replenish volatile elements (sulfur, oxygen, carbon, hydrogen) in situ will alleviate strict leak tolerances for the processing equipment. This makes Mars and C-type asteroids more forgiving in terms of volatile loss, while the Moon and S-type asteroids are relatively volatile poor and therefore less forgiving. Striking a balance between chamber pressure and heating in Reaction 8 will also influence energy efficiency and volatile loss. As Fig. 2 shows, decreasing overbearing pressure during sulfate decomposition will also decrease the temperatures required to drive off the sulfur-bearing gases, but will also require a more robust and energy intensive vacuum system. Validation experiments should explore this balance to determine which pressure and temperature profiles are most efficient to fully decompose the sulfates.

## 3.3 Comparison to other ISRU Methodologies

Although we cannot accurately calculate the energy requirements of the SSAP, we can qualitatively compare its inherent strengths and weaknesses to a collection of other ISRU techniques: molten salt electrolysis (modified FCC Cambridge Process), vapor phase pyrolysis, and hydrogen reduction. An inherent advantage that all of these methods have over the SSAP, is their experimental characterization.

Molten salt electrolysis [52,53], specifically one modified for lunar surface operations [54] can essentially reduce all oxides and silicates into metallic alloys, while also releasing nearly all the oxygen present. This benefit comes at the cost of the required operating temperature. While the SSAP will need to reach similar operating temperatures as this approach (~950 °C), this is only the peak temperature required, while salt electrolysis will need to maintain this temperature for the duration of the process. Special consideration will also need to be taken when choosing which salt and anode to use such that they are not depleted or corroded respectively, though this concern is relatively minor.

Like salt electrolysis, vapor phase pyrolysis [55,56,57] can also reduce the native minerals into metals while also liberating all the oxygen. A major benefit of pyrolysis is its relative simplicity: heating up the rocks they vaporize. This heating however is also the major drawback, since the temperatures required are in excess of 2000 °C.

Another approach for reducing minerals into metals, while also harvesting oxygen involves using hydrogen [34] on the bulk material. This process works mostly on ilmenite, reducing it into titanium oxide and iron metal, while releasing some oxygen (~5 wt% [58]) at the same time. Unfortunately this approach has a lesser effect on silicates. This approach also requires operating temperatures which are slightly higher than the SSAP's peak temperature. An advantage to this approach is its simplicity compared to the SSAP and could conceivably be performed as a pre-processing step for the SSAP.

## 4. Conclusions

In this paper, we have presented the physical and chemical pathways for the proposed Silicate-Sulfuric Acid Process, which aims to manufacture building materials from abundant resources found in situ on major planetary bodies such as the Moon, Mars and asteroids. Although this approach has not yet been tested, it allows for inherent recycling of volatile elements, such that little to no material must be supplied after the initial investment. This proposed ISRU approach could provide substantial building materials including iron metal for 3D printing, silica for window construction, as well as modest amounts of oxygen gas. The next step for further investigating the utility of the SSAP lies in experimental validation.

## 5. Acknowledgments

We would like to thank our anonymous reviewers for their insightful suggestions that considerably improved the scope and quality of this manuscript. This work was funded by the Commonwealth Scientific and Industrial Research Organisation's Top Up Scholarship (No. 50077118).